\documentclass[prl,twocolumn,superscriptaddress]{revtex4-1}
\usepackage{amssymb}
\usepackage{amsfonts}
\usepackage{bbm}
\usepackage{mathrsfs}
\usepackage{graphics,graphicx,epsfig,bm,amsmath,amsthm,amssymb}
\usepackage{bm}
\usepackage{bbm}
\usepackage{longtable}
\usepackage{multirow}
\usepackage{array}
\usepackage{color}
\usepackage{cases}
\usepackage{booktabs }
\usepackage[usenames,dvipsnames]{xcolor}

\usepackage{xcolor}

\usepackage{float}
\usepackage[a4paper,colorlinks=true,
linkcolor=blue,citecolor=blue,
pdfauthor={ },
pdftitle={ },
pdfsubject={ },
pdfkeywords={ }]{hyperref}

\newcommand{\rmnum}[1]{\romannumeral #1}
\newcommand{\Rmnum}[1]{\expandafter\@slowromancap\romannumeral #1@}

\begin{document}

\title{Experimental dissipative quantum sensing}

\author{Yijin Xie}
\affiliation{Hefei National Laboratory for Physical Sciences at the Microscale and Department of Modern Physics, University of Science and Technology of China, Hefei 230026, China}
\affiliation{CAS Key Laboratory of Microscale Magnetic Resonance, University of Science and Technology of China, Hefei 230026, China}
\affiliation{Synergetic Innovation Center of Quantum Information and Quantum Physics, University of Science and Technology of China, Hefei 230026, China}

\author{Jianpei Geng}
\affiliation{3. Physikalisches Institut, University of Stuttgart, Pfaffenwaldring 57, 70569 Stuttgart, Germany}

\author{Huiyao Yu}
\affiliation{Hefei National Laboratory for Physical Sciences at the Microscale and Department of Modern Physics, University of Science and Technology of China, Hefei 230026, China}
\affiliation{CAS Key Laboratory of Microscale Magnetic Resonance, University of Science and Technology of China, Hefei 230026, China}
\affiliation{Synergetic Innovation Center of Quantum Information and Quantum Physics, University of Science and Technology of China, Hefei 230026, China}

\author{Xing Rong}
\email{xrong@ustc.edu.cn}
\affiliation{Hefei National Laboratory for Physical Sciences at the Microscale and Department of Modern Physics, University of Science and Technology of China, Hefei 230026, China}
\affiliation{CAS Key Laboratory of Microscale Magnetic Resonance, University of Science and Technology of China, Hefei 230026, China}
\affiliation{Synergetic Innovation Center of Quantum Information and Quantum Physics, University of Science and Technology of China, Hefei 230026, China}

\author{Ya Wang}
\affiliation{Hefei National Laboratory for Physical Sciences at the Microscale and Department of Modern Physics, University of Science and Technology of China, Hefei 230026, China}
\affiliation{CAS Key Laboratory of Microscale Magnetic Resonance, University of Science and Technology of China, Hefei 230026, China}
\affiliation{Synergetic Innovation Center of Quantum Information and Quantum Physics, University of Science and Technology of China, Hefei 230026, China}

\author{Jiangfeng Du}
\email{djf@ustc.edu.cn}
\affiliation{Hefei National Laboratory for Physical Sciences at the Microscale and Department of Modern Physics, University of Science and Technology of China, Hefei 230026, China}
\affiliation{CAS Key Laboratory of Microscale Magnetic Resonance, University of Science and Technology of China, Hefei 230026, China}
\affiliation{Synergetic Innovation Center of Quantum Information and Quantum Physics, University of Science and Technology of China, Hefei 230026, China}



\begin{abstract}
Quantum sensing utilizes quantum systems as sensors to capture weak signal, and provides new opportunities in nowadays science and technology.
 The strongest adversary in quantum sensing is decoherence due to the coupling between the sensor and the environment. The dissipation will destroy the quantum coherence and reduce the performance of quantum sensing. Here we show that quantum sensing can be realized by engineering the steady-state of the quantum sensor under dissipation. We demonstrate this protocol with a magnetometer based on ensemble Nitrogen-Vacancy centers in diamond, while neither high-quality initialization/readout of the sensor nor sophisticated dynamical decoupling sequences is required.
 Thus our method provides a concise and decoherence-resistant fashion of quantum sensing. The frequency resolution and precision of our magnetometer are far beyond the coherence time of the sensor. Furthermore, we show that the dissipation can be engineered to improve the performance of our quantum sensing.
 By increasing the laser pumping, magnetic signal in a broad audio-frequency band from DC up to $140~$kHz can be tackled by our method.
 Besides the potential application in magnetic sensing and imaging within microscopic scale, our results may provide new insight for improvement of a variety of high-precision spectroscopies based on other quantum sensors.
\end{abstract}

\maketitle

Quantum sensing utilizes a quantum system to perform a measurement of a physical quantity, and it allows one to gain advantages over its classical counterpart \cite{RevModPhys.89.035002}.  There are variety of quantum systems for quantum sensing, such as neutral atoms \cite{NeutralAtoms}, trapped ions\cite{Maiwald:2009aa,Biercuk:2010aa}, solid-state spins \cite{Taylor2008}, and superconducting circuits \cite{RevModPhys.89.035002}.
The regular procedure of quantum sensing includes three elementary steps:
($\rmnum{1}$) Initialize the state of the sensor to a superposition state.
($\rmnum{2}$) Let the sensor interact with the target field.
($\rmnum{3}$) Readout the final state of the  sensor.
For the step ($\rmnum{1}$) and ($\rmnum{3}$), high-quality and efficient  initialization and readout are required.
The step ($\rmnum{2}$) is usually very fragile, due to the inevitable interactions between the sensor and the environment. To minimize the effect of the environmental noise, exquisite dynamical decoupling technologies have been developed to suppress the noise \cite{deLange60} and enhance the performance of quantum sensing \cite{Taylor2008}.
However, the decoherence sets an upper bound on the time delay between the pulses in the dynamical decoupling sequence, corresponding to a lower bound of the detectable frequency of signal. As a result, the detection of low frequency magnetic field, which is important in magnetic navigation \cite{6432396,56910}, magnetic-anomaly detection \cite{56910}  and bio-magnetic field detection \cite{Wikswo53, RevModPhys.65.413, Barry14133}, can hardly benefit from the dynamical decoupling protocols.

\begin{figure}
\centering
\includegraphics[width=\columnwidth]{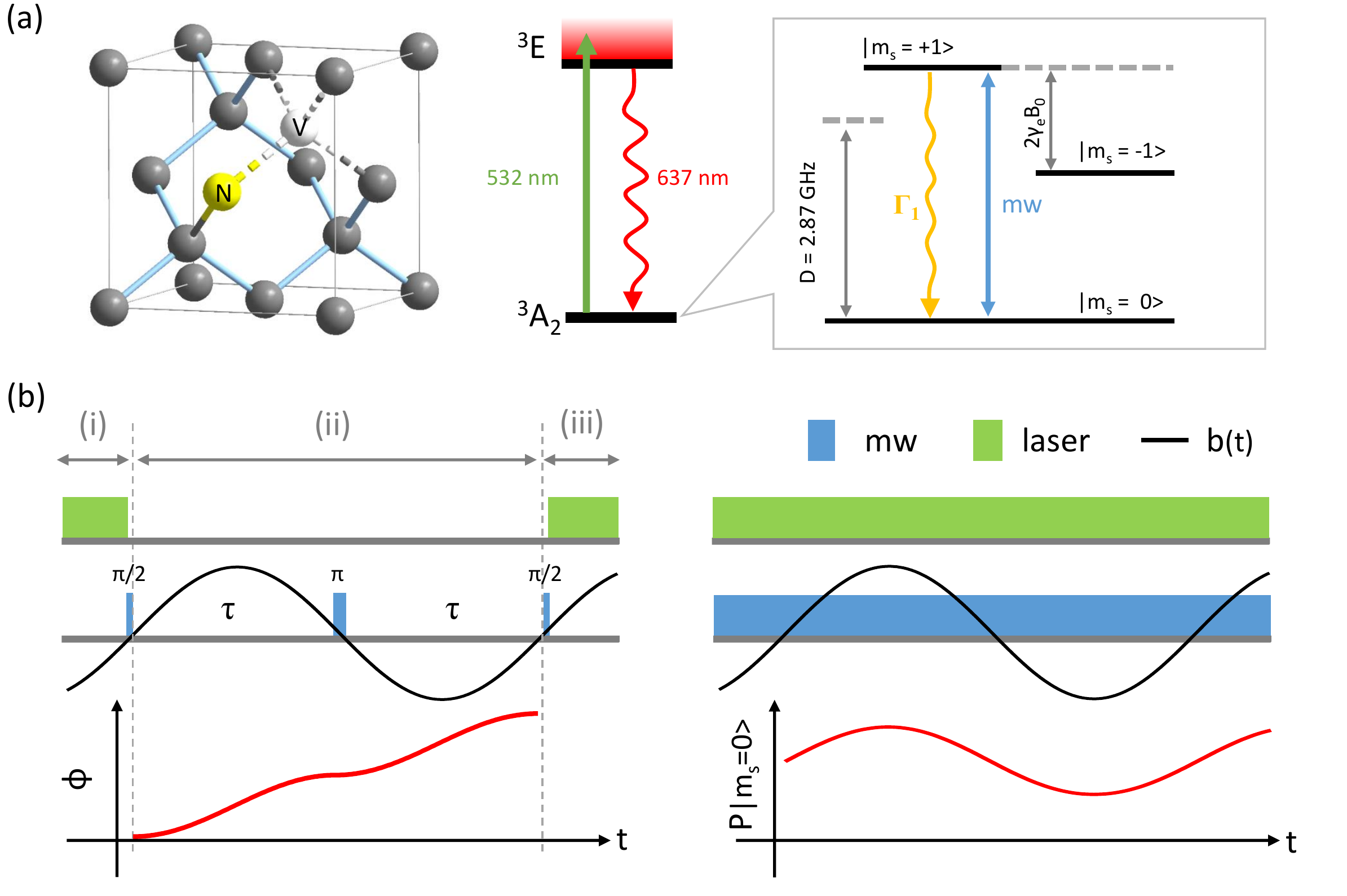}
\caption{Two types of quantum sensing based on NV centers.  (\textbf{a}) Left: the diamond crystal lattice with an NV center defect.
Right: the energy level diagram of NV center. The ground and excited states are denoted as $^3A_2$ and $^3E$. The spin triplet $^3A_2$ contains three spin sublevels $|m_s=0\rangle$ and $|m_s=\pm1\rangle$. The state $|m_s=0\rangle$ and $|m_s=+1\rangle$ are encoded as the quantum sensor. The state of the quantum sensor can be initialized and read out with 532 nm laser pulses, and manipulated with microwave pulses. The initialization with laser pulse can be considered as a dissipative process in which the population decays from $|m_s=+1\rangle$ to $|m_s=0\rangle$ with rate $\Gamma_1$. (\textbf{b}) Left: regular procedure of quantum sensing.
Bottom: the accumulated relative phase shift $\phi$ is plotted as function of time $t$.  Right: quantum sensing with steady-states. Laser and microwave fields have been continuously applied on the NV center. Bottom: the dynamics of steady state can be continues monitored by measuring the population of state $|m_s=0\rangle$.
\label{fig1}}
\end{figure}

Inspired by resent progress in dissipative quantum computation and quantum metrology \cite{Verstraete2009,Reiter2017,Lin2013},
we propose and experimentally demonstrate a dissipative quantum sensing protocol for low frequency signal detection in an important type of solid-state quantum sensors, Nitrogen-Vacancy (NV) center in diamond \cite{DOHERTY20131}. Due to the atomic scale of NV centers, quantum sensing based on NV center provides exciting quantum technologies, such as nuclear magnetic resonance and magnetic resonance imaging with nanoscale.
An ensemble electron spins of NV center is taken as a quantum sensor for magnetic field measurement.
The steady state of the sensor under dissipation can be engineered to be sensitive to the detected magnetic field.
We show that the frequency resolution and precision go far beyond the spin coherence time.
Furthermore, the laser pumping procedure, during which the sensor can be initialized and read out, can be utilized to introduce an additional dissipation to the sensor.
We also show that such a laser-controlled dissipation can improve detection bandwidth for quantum sensing.
Our method is capable to sense magnetic signal over a broad audio-frequency band ranging from DC to $140~$kHz.

Figure \ref{fig1}(a) shows the schematic of the NV center, which is an atomic defect consisting of a substitutional nitrogen and a vacancy adjacent to it.
It is negatively charged since the center comprises of six electrons, two of which are unpaired.
The energy level diagram is shown in the right panel of Figure \ref{fig1}(a).
The electronic ground state is a spin triplet state $^3A_2$ consisting of three spin sublevels $|m_s=0\rangle$ and $|m_s=\pm1\rangle$.
The NV center can be excited from $^3A_2$ to the excited state $^3E$ by a laser pulse with $532~$nm, and decays back to $^3A_2$ emitting photoluminescence.
The photoluminescence intensity is dependent on the spin state of the NV center, which can be utilized for readout of the spin state.
Due to the intersystem crossing process, there is a larger probability to decay into $|m_s=0\rangle$ of $^3A_2$ than $|m_s=\pm1\rangle$.
Therefore, there is approximately a net decay rate $\Gamma_1$ from $|m_s=+1\rangle$ (or $|m_s=-1\rangle$) to $|m_s=0\rangle$ of $^3A_2$ after a round of the optical transition.
The decay rate $\Gamma_1$ can be harnessed with controllable power of the laser pulse.
The optical transitions can be utilized for initialization of the spin state.
The spin states $|m_s=0\rangle$ and $|m_s= +1\rangle$ of $^3A_2$ are encoded as a quantum sensor.
This sensor can be manipulated by microwave field with angular frequency being $\omega_e = D - \gamma_e B_0$, where $D =2\pi\times 2.87~$GHz is the ground state zero splitting, $\gamma_e = -2\pi\times28~$GHz/T is the gyromagnetic ratio of the electron spin and $B_0$ is the external magnetic field.

Left panel of figure \ref{fig1}(b) presents the regular quantum sensing based on NV center.
In part ($\rmnum{1}$), a laser pulse together with a $\pi/2$ microwave pulse are applied on the NV center to prepare the quantum sensor to the superposition state $(|0\rangle-i|1\rangle)/\sqrt{2}$.
In the procedure labeled by part ($\rmnum{2}$), we let the quantum sensor interact with the magnetic signal.
The magnetic signal contributes to a relative phase shift $\phi$ of the quantum sensor state.
A spin echo technique is utilized to prolong the coherence time of the sensor.
A $\pi$ pulse is applied to decouple the interaction between the quantum sensor and the environmental noise.
If the magnetic signal is in phase with the spin echo pulse sequence, the relative phase shift $\phi$ due to the magnetic signal is accumulated rather than cancelled as shown in the bottom of left panel of figure \ref{fig1}(b).
More sophisticated dynamical decoupling (DD) sequences can be applied instead of spin echo sequence to enhance the coherence time of the NV center.
Once the coherence time is prolonged, the accumulated phase shift $\phi$ increases and the sensitivity of quantum sensing will be improved \cite{PhysRevB.86.045214}.
The information of $\phi$ can be extracted by procedure ($\rmnum{3}$) consisting of a $\pi/2$ microwave pulse and a laser pulse.
The lowest frequency of detected signal is bounded by $1/2\tau$ \cite{Boss837}, where $\tau$ stands for the delay time between microwave pulses.
However, this delay time is limited by spin docoherence, so that low frequency signal detection by this method is challenging. Furthermore, once multiple-pulse DD sequences are applied, imperfection of DD pulses contributes to the reduction of sensitivity.
The non-ideal initialization and readout of the NV center also contribute to the reduction of observed signal.

Right panel of figure \ref{fig1}(b) shows the basic idea of our dissipative quantum sensing protocol.
Laser and microwave fields are always turned on simultaneously during the quantum sensing.
The laser pumping introduces an additional dissipation with decay rate $\Gamma_1$ from $|m_s=+1\rangle$ to $|m_s=0\rangle$.
The dissipation can be described by an amplitude damping process.
The microwave field drives the electron spin continuously, corresponding to the evolution of the sensor governed by the Hamiltonian $H=\Delta S_z - \gamma_e b(t)S_z - \gamma_e B_1 S_x$.
Here the detuning $\Delta \equiv \omega_e - \omega_{mw}$ is the difference between the transition angular frequency of the electron spin $\omega_e$ and the angular frequency of the microwave field $\omega_{mw}$, $B_1$ corresponds to the strength of the microwave field, $b(t)$ stands for the ac magnetic field to be detected along the NV axis, and $S_{x, y, z}=\sigma_{x, y, z}/2$ are the spin operators with $\sigma_{x, y, z}$ being the Pauli operators.
The ac magnetic field can be written as $b(t ) = b_{ac}\cos(\omega_{ac} t + \phi_{ac})$, where $b_{ac}$, $\omega_{ac}$ and $\phi_{ac}$ are the amplitude, angular frequency, and phase, respectively.
Besides the dissipation introduced by the laser pumping, the sensor undergoes dephasing due to the interaction with environment (e.g. the nuclear spin bath).
The dephasing, with a dephasing rate $\Gamma_2$, can be described by a phase damping process.
The master equation, which describes the dynamics of the state $\rho$ of the sensor in a rotating frame, can be written as

\begin{equation}
\frac{\mathrm{d}\rho}{\mathrm{d} t} = -i [H, \rho] + \sum_{j =1,2} (2 L_j\rho L_j^\dag - L_j^\dag L_j\rho -\rho L_j^\dag L_j).
\label{master equation}
\end{equation}
The first term on the right-hand side of equation \ref{master equation} stands for the evolution under the Hamiltonian $H$.
The second term on the right-hand side of equation \ref{master equation} describes the dissipation process of the system.
The operator $L_1 = \sqrt{\Gamma_1/2}\sigma_-$ corresponds to the amplitude damping, and $L_2 = \sqrt{\Gamma_2}/2\sigma_z$ corresponds to the phase damping, where $\sigma_- = (\sigma_x - i\sigma_y)/2$.
Here, the intrinsic longitudinal relaxation of the sensor is neglected, since the intrinsic relaxation time (about $3.9~$ms) is much longer than the decay time under the laser pumping ($1/\Gamma_1$, about several microseconds).
The dissipation process leads the sensor to a steady state.
The characteristic time to reach the steady state depends on two parameters: the dephasing time under the laser, $T_2=1/(\Gamma_2+\Gamma_1/2)$, which is measured to be about $200~$ns, and the decay time of amplitude damping, $T_1 = 1/ \Gamma_1$.

For low frequency signal detection, the ac magnetic field to detect varies in a way that is much slower than that the sensor reaches a steady state.
In this case, the ac magnetic field can be considered quasi-static, and the state of the sensor can be approximated by the steady state under the quasi-static magnetic field.
%
For weak signal detection, i.e., if $|\gamma_e b(t)T_2|\ll1$, the steady state $\rho$ can be approximated, up to the first order, as
\begin{equation}
\label{steady-state-1stOrder}
\rho=\rho_0+Kb(t),
\end{equation}
where $\rho_0$ and $Kb(t)$ stand for the time independent and dependengt part of the steady state, respectively.
The detailed information of $\rho_0$ and K are provided in the Supplemental Material \cite{Supp}.

The ac magnetic field to detect is encoded into the state of the quantum sensor in a linear fashion according to equation \ref{steady-state-1stOrder}.
Since the laser is always applied on the NV centers, the photoluminescence intensity which reflects the probability of the state in $|m_s = 0\rangle$ can be continuously monitored.
The probability in $|m_s = 0\rangle$ is
\begin{equation}
\label{P0}
\begin{aligned}
P_{|m_s=0\rangle}=\frac{2+s+2\Delta^2T_2^2}{2(1 + s +\Delta^2T_2^2) }-\frac{\gamma_e\Delta T_2^2s}{(1+s+\Delta^2T_2^2)^2}b(t),
\end{aligned}
\end{equation}
where $s = \gamma_e^2B_1^2 T_1 T_2$. The dynamics of the probability $P_{|m_s = 0\rangle}$  is plotted schematically in the bottom of right panel of figure \ref{fig1}(b).
The oscillation of  $P_{|m_s = 0\rangle}$ reflects the amplitude and frequency of the ac magnetic field $b(t)$.
When the detuning is set to $\Delta = \sqrt{1+ s}/\sqrt{3}T_2 $, the optimal sensitivity for signal detection is reached.
Since $\rho$ is a steady state which can last for arbitrary long time, the detection can be, in principle, in an arbitrary precision.


The experimental setup has been developed for ensemble NV based magnetometer. Always-on laser and microwave fields are applied on NV centers. The microwave field is fed on the NV centers by a double split-ring resonator. A coil is used to send the magnetic field $b(t)$ on NV centers. The waveform of $b(t)$  can be generated by the wave generator (33500B, Keysight). Two photodiodes (PDs) are used to transform intensities of green laser and red fluorescence to voltages, which can be monitored by a two channel lock-in amplifier (HF2LI, Zurich Instruments). The time constant of the lockin-amplifier is set to zero when it is utilized as an oscilloscope. The voltage, $v_1$, from PD1, which receives the red fluorescence, is proportional to the probability $P_{|m_s=0\rangle}$ of state in $|m_s = 0\rangle$.  The voltage, $v_2$, from PD2, which receives the green laser, is used to cancel the long time drift due to the laser power instability.

\begin{figure}
\includegraphics[width=\columnwidth]{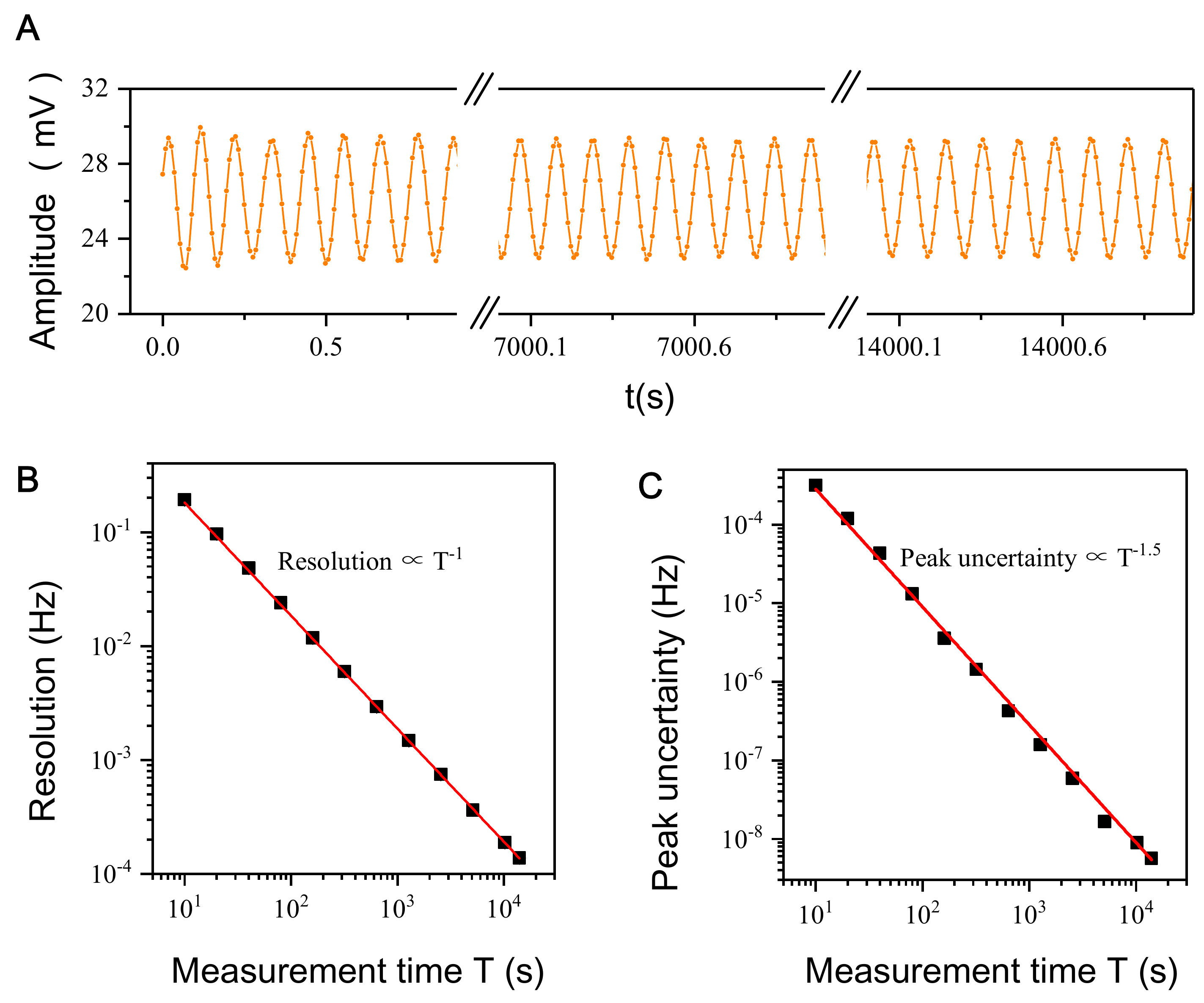}
\caption{Experimental frequency resolution and precision measurement.   (\textbf{a}) Experimental data for detection of a $9~$Hz magnetic field. The voltage reflects the population in $|m_s=0\rangle$ of the sensor state. A coherent oscillation which lasts for about four hours without decay is observed, showing that the sensor is in a steady state.
(\textbf{b}) Frequency resolution as a function of total measurement time $T$. Solid line indicates that the resolution improves with measurement time $T$ as $ T^{-1}$.
 (\textbf{c}) Precision of frequency estimation as a function of measurement time $T$. Solid line indicates that the precision is improved as $ T^{-3/2}$.
\label{fig2}}
\end{figure}

Figure \ref{fig2}a shows the experimental data of $v_1$, which reflects $P_{|m_s=0\rangle}$ of the steady state, as a function of time when the frequency of $b(t)$ is set to 9 Hz.
The voltage $v_1$ is recorded by the lock-in amplifier which is utilized as an oscilloscope with the frequency of its reference signal set to zero.
It can be observed that there is a coherent oscillation which lasts for about four hours without decay.
The non-decay behavior shows that the quantum state of the sensor is a steady state as expected.
The ac magnetic field $b(t)$ is encoded into $v_1$ according to equation \ref{P0}.
The frequency of $b(t)$ is the same as that of $v_1$, and the amplitude of $b(t)$ is proportional to that of $v_1$.
The fast Fourier transformation of $v_1$ provides the spectrum of $b(t)$ in the frequency domain.
The peak position of the spectrum corresponds to the frequency of $b(t)$, and the linewidth of the spectrum is defined as the frequency resolution.
The frequency and frequency resolution are obtained by fitting the spectrum, with the fitting uncertainty of the frequency defined as the frequency precision \cite{Boss837,Schmitt832}.
Figure \ref{fig2}b shows the obtained frequency resolution as a function of the measurement time $T$.
The frequency resolution improves with the measurement time as $T^{-1}$.
When the data of $v_1$ is measured for $T=14000~$s, a frequency resolution of $138~\mu$Hz is obtained.
Figure \ref{fig2}c shows the frequency precision as a function of the measurement time $T$.
The frequency precision improves with the measurement time as $T^{-3/2}$.

\begin{figure}
\centering
\includegraphics[width=\columnwidth]{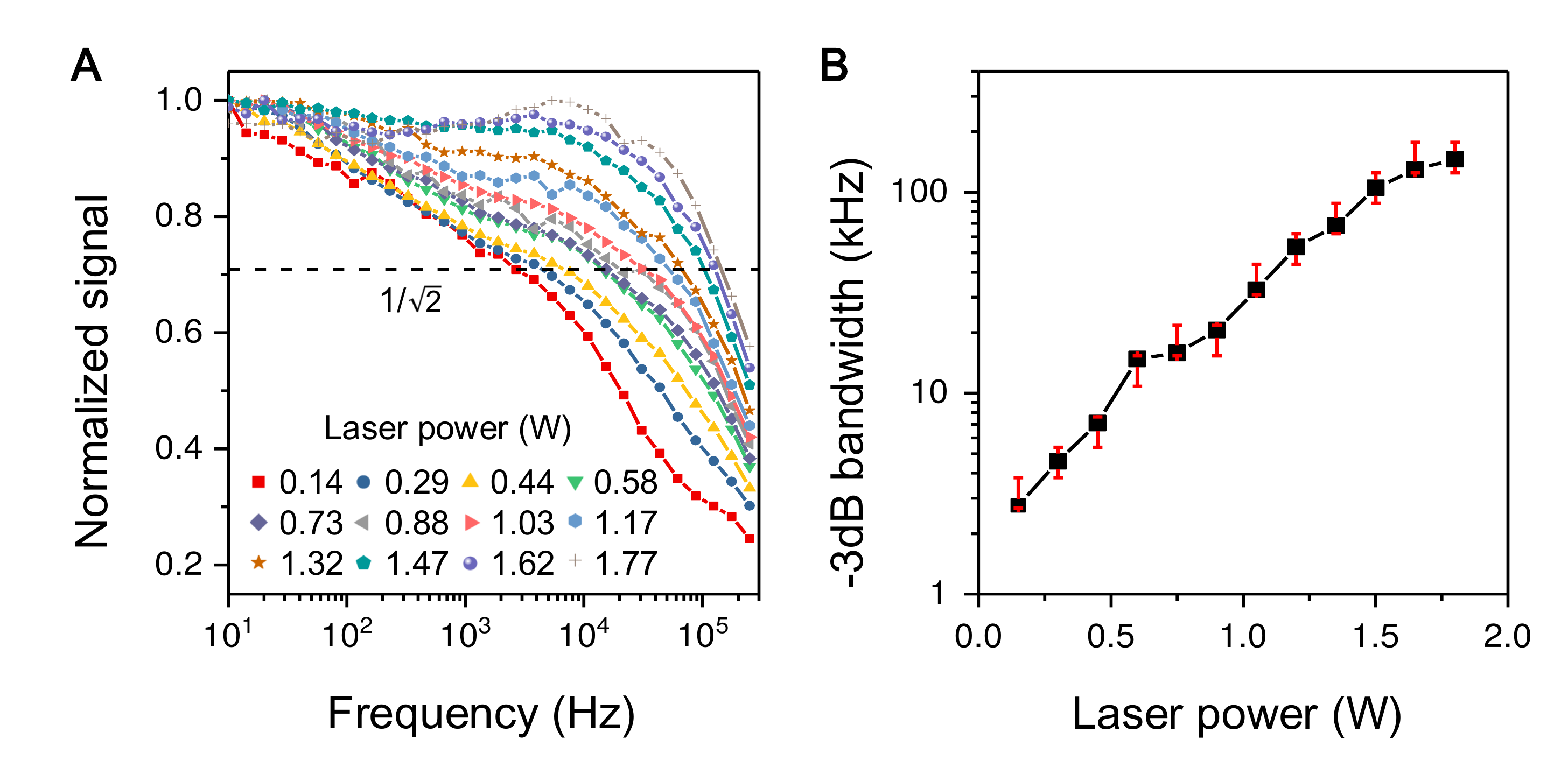}
\caption{Experimental measurement of the detection bandwidth. (\textbf{a}) The normalized measured amplitudes of $v_1$ as functions of the frequency of $b(t)$ with different values of laser power. The detection bandwidth is defined as the frequency of $b(t)$ at which the normalized amplitude decreases to $1/\sqrt{2}$. (\textbf{b}) The experimental detection bandwidth as a function of laser power.
\label{fig3}}
\end{figure}

Figure \ref{fig3} shows the experimental measurement of the detection bandwidth of our dissipative quantum sensing protocol.
Since the protocol is based on the steady state under a quasi-static magnetic field $b(t)$, the detection bandwidth of $b(t)$ depends on the rate that the sensor reaches the steady state.
The rate to reach the steady state can be engineered by varying the laser power, since the laser power influences the decay time $T_1$ of the amplitude damping.
At a certain laser power, as the frequency of $b(t)$ increases, the rate that $b(t)$ varies will be more and more comparable with the rate that the sensor reaches the steady state.
If the frequency of $b(t)$ reaches a value large enough, the assumption of quasi-static $b(t)$ will be broken down and the quantum state of the sensor will be no longer a steady state.
Therefore, the amplitude of $v_1$ is expected to decrease with increasing frequency of $b(t)$.
In the experiment, the amplitude of $v_1$ is measured by the lock-in amplifier with the frequency of its reference signal set to that of $b(t)$.
Figure \ref{fig3}a shows the measured amplitudes as functions of the frequency of $b(t)$ when the laser power is set to a series of values.
For clarity, the amplitudes are normalized so that their values at the first point equal 1.
As expected, the normalized amplitude decreases as the frequency of $b(t)$ increases for any certain laser power.
The frequency at which the normalized amplitude decreases to $1/\sqrt{2}$ is defined as the detection bandwidth.
Figure \ref{fig3}b shows the detection bandwidth as a function of the laser power.
It clearly shows that the detection bandwidth increases with the laser power increasing.
When the laser power is set to $1.8~$W, the detection bandwidth of $146~$kHz is reached.
Thus the improvement of the detection bandwidth by engineering the laser power has been demonstrated.

In conclusion, we have proposed and experimentally demonstrated a dissipative quantum sensing protocol based on the steady state of the sensor of ensemble NV centers in diamond.
The frequency resolution and precision is far beyond the limit of the sensor's coherence time.
We experimentally show that the frequency resolution and precision can be improved with the measurement time $T$ as $T^{-1}$ and $T^{-3/2}$, respectively.
The detection bandwidth in our protocol can be engineered by controlling dissipation.
As has been demonstrated in the experiment, the detection bandwidth increases with the laser power increasing.
When the laser power is set to $1.8~$W, the detection of ac magnetic field over a broad band ranging from DC to about $140~$kHz can be achieved. Our method is essentially different from magnetic field detection by continuous-wave method based on NV centers\cite{PhysRevApplied.10.034044,Barry14133}, whose bandwidth is limited by the time constant of the lockin-amplifier.
The sensitivity of our setup is measured to be about $1~$nT/$\sqrt{Hz}$ when the frequency is higher than 1 kHz \cite{Supp}.
The sensitivity can be further improved with optimization of the experimental parameters of our apparatus (e.g. the material properties of the diamond).
The dynamic range of our protocol is measured to be better than 80 dB when the frequency is larger than 1 kHz \cite{Supp}.
The high dynamic range can be achieved with our method, because we can continuously monitor the quantum probe and get rid of the disadvantage of  regular procedure of quantum sensing \cite{Waldherr2011}.
Our protocol provides new possibility and insight in quantum sensing, imaging, and spectroscopies based on quantum sensors.

This work was supported by the National Key R$\&$D Program of China (Grants No. 2018YFA0306600 and No. 2016YFB0501603), the CAS (Grants No. GJJSTD20170001, No.QYZDY-SSW-SLH004 and No.QYZDB-SSW-SLH005), the NNSFC (Grants No. 81788101, No. 11761131011), and Anhui Initiative in Quantum Information Technologies (Grant No. AHY050000). X.R. thanks the Youth Innovation Promotion Association of Chinese Academy of Sciences for the support.

Y. X. and  J. G.  contributed equally to this work.

\clearpage
\onecolumngrid
\vspace{1.5cm}
\begin{center}
\textbf{\large Supplementary material for "Experimental dissipative quantum sensing"}
\end{center}
\setcounter{figure}{0}
\setcounter{equation}{0}
\setcounter{table}{0}
\makeatletter
\renewcommand{\thefigure}{S\arabic{figure}}
\renewcommand{\theequation}{S\arabic{equation}}
\renewcommand{\thetable}{S\arabic{table}}
\renewcommand{\bibnumfmt}[1]{[RefS#1]}
\renewcommand{\citenumfont}[1]{RefS#1}

\section{Hamiltonian of the quantum probe}
The ensemble electron spins of nitrogen-vacancy centers in diamond are utilized as the quantum probe for detection of low frequency ac magnetic field.
The Hamiltonian of the NV center is described as
\begin{equation}
\label{H_tot}
H_{\textrm{NV}}=H_{\textrm{ZF}}+H_{\textrm{ZM}}+H_{\textrm{MW}}+H_{\textrm{d}}+H_{\textrm{hyp}}.
\end{equation}
The first term, $H_{\textrm{ZF}}=DS_{z,3}^2$, corresponds to the zero-field splitting of the NV center electron spin, with $D=2\pi\times2.87~$GHz and $S_{j,3}~(j=x, y, z)$ being the electron spin operator.
The subscript "3" denotes that the NV center electron spin is a 3-level (spin-1) system.
The second term, $H_{\textrm{ZM}}=-\gamma_eB_0S_{z,3}$, is the Zeeman splitting of the electron spin under a static magnetic field $B_0$ along the NV axis, where $\gamma_e=-2\pi\times28~$GHz/T is the gyromagnetic ratio.
The third term, $H_{\textrm{MW}}=-\gamma_e\times\sqrt{2}B_1\cos\omega_{mw} tS_{x,3}$, is the control Hamiltonian introduced by a microwave field with amplitude $\sqrt{2}B_1$ and angular frequency $\omega_{mw}$.
The fourth term, $H_\textrm{d}=-\gamma_eb(t)S_{z,3}$, is the term for detection with $b(t)=b_{ac}\cos(\omega_{ac}t+\phi_{ac})$ being the ac magnetic field to be detected along the NV axis.
The fifth term, $H_{\textrm{hyp}}=AS_{z,3}I_{z,3}$, is the hyperfine interaction with the $^{14}$N nuclear spin, with $A=-2\pi\times2.16~$MHz being the hyperfine coupling strength and $I_{z,3}$ being the $^{14}$N nuclear spin operator.
In the experiment, the $^{14}$N nuclear spin is not polarized.
Dependent on the state $|m_I\rangle$ of the nuclear spin, the Hamiltonian of the electron spin can be simplified into
\begin{equation}
\label{H_mI}
H_{m_I}=DS_{z,3}^2+(-\gamma_eB_0+m_IA)S_{z,3}-\sqrt{2}\gamma_eB_1\cos\omega_{mw} tS_{x,3}-\gamma_eb(t)S_{z,3},
\end{equation}
where $m_I=-1, 0, 1$.

The frequency of the microwave field, $\omega_{mw}/2\pi$, is set close to the transition frequency $\omega_e/2\pi$ between $|m_s=0\rangle$ and $|m_s=1\rangle$, where $\omega_e=D-\gamma_eB_0+m_IA$.
The transition probability between $|m_s=0\rangle$ and $|m_s=-1\rangle$ is negligible due to the large detuning.
Considering the microwave field together with the always-on laser field which initializes the NV center electron spin into $|m_s=0\rangle$, there is negligible population on state $|m_s=-1\rangle$ during the experiment.
Therefore, the subspace spanned by $|m_s=1\rangle$ and $|m_s=0\rangle$ is considered and the Hamiltonian of the quantum probe is written as
\begin{equation}
\label{H_probe}
H_\textrm{probe}=\omega_eS_z-2\gamma_eB_1\cos\omega_{mw} tS_x-\gamma_eb(t)S_z,
\end{equation}
where $S_j=\sigma_j/2~(j=x,y,z)$ is the spin operator of the equivalent two-level system with Pauli operator $\sigma_z=|m_s=1\rangle\langle m_s=1|-|m_s=0\rangle\langle m_s=0|$, $\sigma_x=|m_s=1\rangle\langle m_s=0|+|m_s=0\rangle\langle m_s=1|$, and $\sigma_y=-i|m_s=1\rangle\langle m_s=0|+i|m_s=0\rangle\langle m_s=1|$.

By turning into a rotating frame which rotates along z-axis with angular frequency $\omega_{mw}$ relative to the laboratory frame, and by considering the rotating-wave approximation, we present the Hamiltonian in the rotating frame as
\begin{equation}
\label{H}
H=\Delta S_z-\gamma_eb(t)S_z-\gamma_eB_1S_x,
\end{equation}
with $\Delta=\omega_e-\omega_{mw}$ being the detuning.

\section{Steady state of the quantum probe}
The state evolution of the quantum probe can be described by the master equation
\begin{equation}
\label{MasterEq}
\frac{\mathrm{d}\rho}{\mathrm{d} t} = -i [H, \rho] + \sum_{j =1,2} (2 L_j\rho L_j^\dag - L_j^\dag L_j\rho -\rho L_j^\dag L_j),
\end{equation}
where $\rho$ is the state of the quantum probe, $L_1=\sqrt{\Gamma_1/2}\sigma_-$ (with $\sigma_-=(\sigma_x-i\sigma_y)/2$) and $L_2=\sqrt{\Gamma_2}/2\sigma_z$ correspond to the amplitude damping and phase damping process.
If the ac magnetic field $b(t)$ to be detected varies sufficiently slowly, $H$ can be considered as a quasi-static Hamiltonian at each time, and the state of the quantum probe can be considered as a steady state slowly varying in step with $b(t)$.
By setting the right hand of Eq. \ref{MasterEq} to zero, the steady state can be derived as
\begin{equation}\label{steady-state}
\rho = \frac{1}{2}\left(
         \begin{array}{cc}
             \frac{s}{1 + s +[\Delta-\gamma_e b(t)]^2T_2^2 } & \frac{\gamma_e B_1 [ \Delta- \gamma_e b(t) ]T_2^2 + i\gamma_e B_1 T_2}{1 + s +[\Delta- \gamma_e b(t)]^2T_2^2 }\\
            \frac{\gamma_e B_1 [ \Delta- \gamma_e b(t)]T_2^2 - i\gamma_e B_1 T_2 }{1 + s +[\Delta- \gamma_e b(t)]^2T_2^2} &   \frac{2 + s + 2[ \Delta-\gamma_e b(t)]^2T_2^2 }{1 + s +[\Delta-\gamma_e b(t)]^2T_2^2 }   \\
         \end{array}
       \right),
\end{equation}
where $T_1=1/\Gamma_1$, $T_2=1/(\Gamma_2+\Gamma_1/2)$, and $s=\gamma_e^2B_1^2 T_1 T_2$.

For weak signal detection, i.e., if $|\gamma_e b(t)T_2|\ll1$, the steady state $\rho$ can be approximated to the first order of $\gamma_e b(t)T_2$, which is
\begin{equation}
\label{steady-state-1stOrder}
\rho=\rho_0+Kb(t),
\end{equation}
with
\begin{equation}
\label{rho0K}
\begin{aligned}
&\rho_0 = \frac{1}{2}\left(
         \begin{array}{cc}
             \frac{s}{1 + s +\Delta^2T_2^2 } & \frac{\gamma_e B_1 \Delta T_2^2 + i\gamma_e B_1 T_2}{1 + s +\Delta^2T_2^2 }\\
            \frac{\gamma_e B_1 \Delta T_2^2 - i\gamma_e B_1 T_2 }{1 + s +\Delta^2T_2^2} &   \frac{2 +s + 2\Delta^2T_2^2 }{1 + s +\Delta^2T_2^2 }   \\
         \end{array}
       \right), \\
&K = -\left(
      \begin{array}{cc}
        -\frac{\gamma_e\Delta T_2^2s}{(1+s+\Delta^2T_2^2)^2} & \frac{\gamma_e^2B_1T_2^2(1+s-\Delta^2T_2^2)-i2\gamma_e^2B_1\Delta T_2^3}{2(1+s+\Delta^2T_2^2)^2} \\
        \frac{\gamma_e^2B_1T_2^2(1+s-\Delta^2T_2^2)+i2\gamma_e^2B_1\Delta T_2^3}{2(1+s+\Delta^2T_2^2)^2} & \frac{\gamma_e\Delta T_2^2s}{(1+s+\Delta^2T_2^2)^2} \\
      \end{array}
    \right).
\end{aligned}
\end{equation}
The probability of the state $\rho$ in $|m_s=0\rangle$ is
\begin{equation}
\label{P0}
P_{|m_s=0\rangle}=\frac{2+s+2\Delta^2T_2^2}{2(1 + s +\Delta^2T_2^2) }-\frac{\gamma_e\Delta T_2^2s}{(1+s+\Delta^2T_2^2)^2}b(t),
\end{equation}

\begin{figure}
\includegraphics[width=0.8\columnwidth]{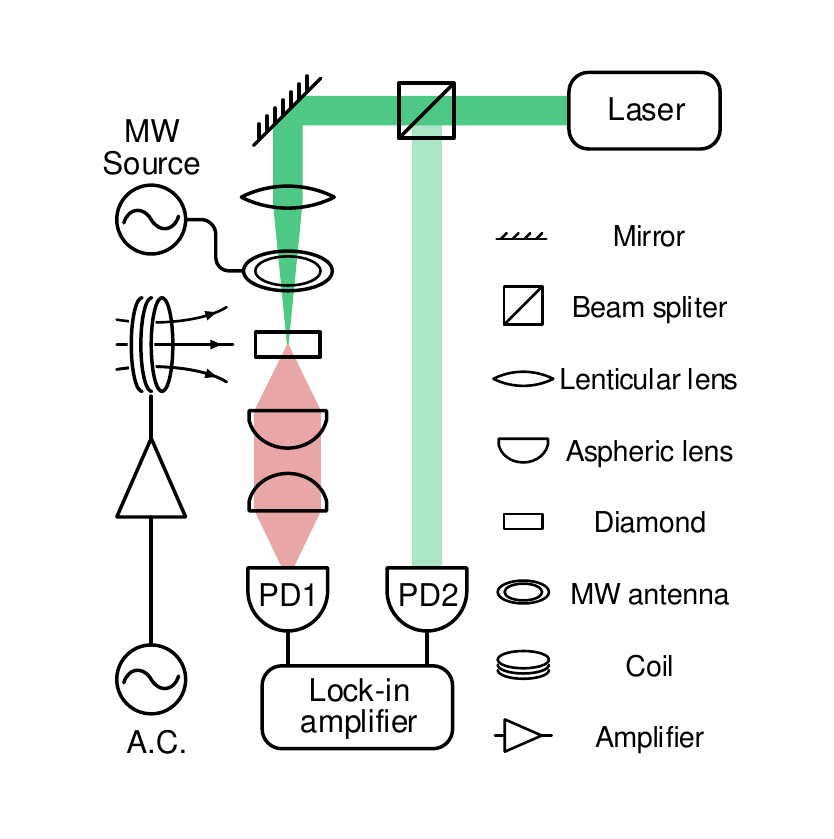}
\caption{\label{ssetup}Schematic of the optically detected magnetic resonance setup for magnetometer based on Nitrogen-Vacancy centers in diamond. Green laser was led to the NV centers. A double-ring resonator was utilized to generate microwave field, $B_1$, on NV centers in a diamond, whose nitrogen concentration is about 10 ppm. The red fluorescence was collected by a photon diode (PD1). The reference laser intensity was detected by another photon diode (PD2). The voltages from the two diodes are sent to a lockin amplifier for further signal precessing. A home-built amplifier was used to amplify the current from the wave generator to the coil, which provided the detected magnetic field, $b(t)$.}
\end{figure}

\section {Schematic of the optically detected magnetic resonance setup }
The experimental setup has been developed for ensemble NV based magnetometer Fig. \ref{ssetup}. Always-on laser and microwave fields are applied on NV centers. The microwave field is fed on the NV centers by a double split-ring resonator. A coil is used to send the magnetic field $b(t)$ on NV centers. The waveform of $b(t)$  can be generated by the wave generator (33500B, Keysight). Two photodiodes (PDs) are used to transform intensities of green laser and red fluorescence to voltages, which can be monitored by a two channel lock-in amplifier (HF2LI, Zurich Instruments). The time constant of the lockin-amplifier is set to zero when it is utilized as an oscilloscope. The voltage, $v_1$, from PD1, which receives the red fluorescence, is proportional to the probability $P_{|m_s=0\rangle}$ of state in $|m_s = 0\rangle$.  The voltage, $v_2$, from PD2, which receives the green laser, is used to cancel the long time drift due to the laser power instability.

\section {Sensitivity and dynamical range}
\begin{figure}[http]
\centering
\includegraphics[width=0.8\columnwidth]{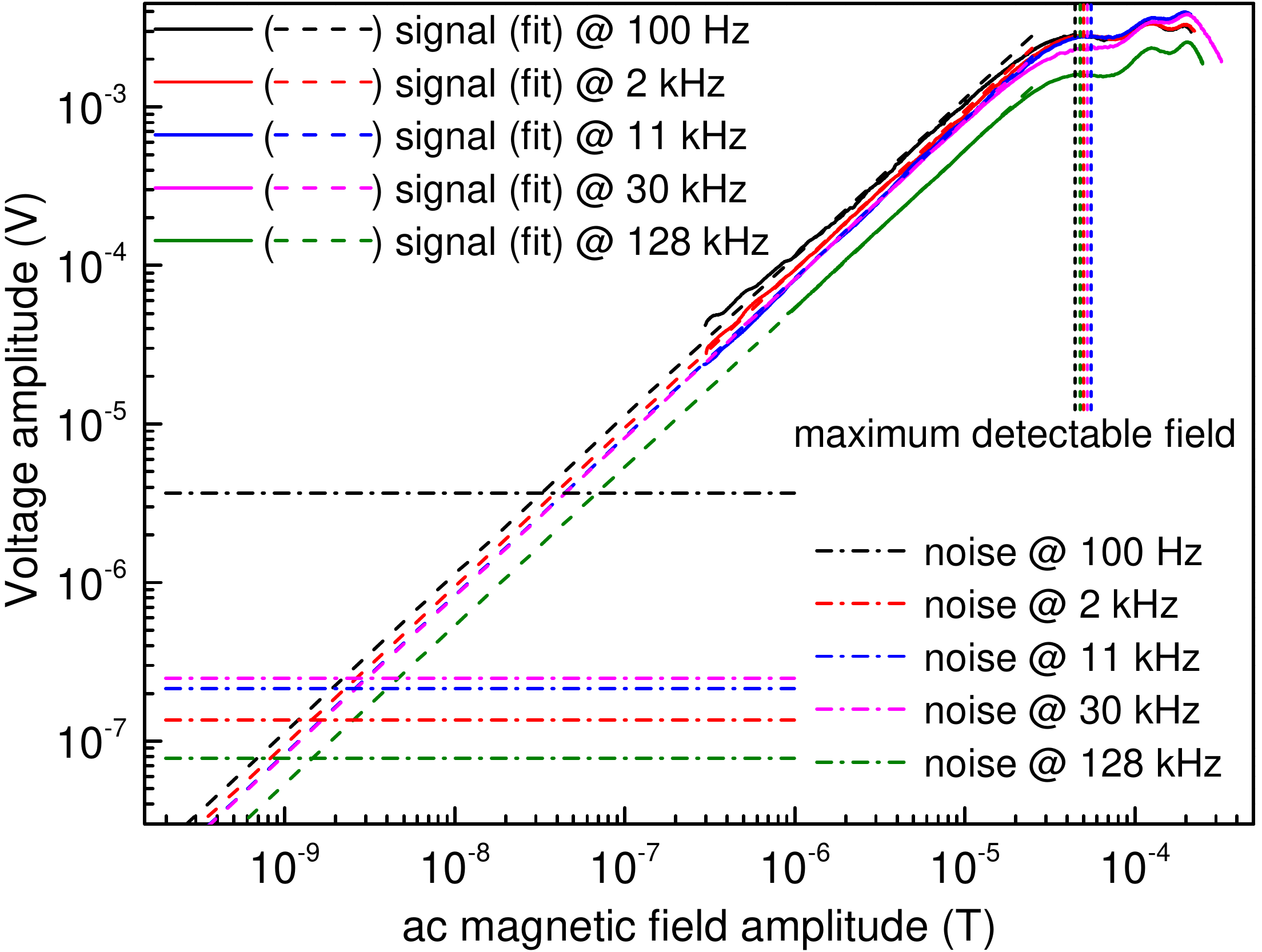}
\caption{\label{FigS1} Measured voltage amplitude $v_{1m}$ of the signal $v_1$, which reflects the probability $P_{|m_s=0\rangle}$ of the state in $|m_s=0\rangle$, as a function of the amplitude $b_{ac}$ of the ac magnetic field to detect. The black, red, blue, magenta, and olive solid lines show the experimentally measured voltage amplitude when the frequency of the ac magnetic field is $100~$Hz, $2~$kHz, $11~$kHz, $30~$kHz, and $128~$kHz, respectively. The dash lines show the fit of the measured voltage amplitude with linear function $v_{1m}=kb_{ac}$, where $k$ is the coefficient. The dash dot lines show the noise level in a bandwidth of $1~$Hz at these frequencies. The short dash lines show the positions of the maximum detectable amplitudes of the magnetic field.}
\end{figure}

For estimation of the sensitivity and dynamical range of our quantum probe in the dissipative sensing protocol, the voltage $v_1$ from the photodiode, which reflects the intensity of the red fluorescence and thus the probability $P_{|m_s=0\rangle}$ of the state in $|m_s=0\rangle$, is measured for various amplitudes and frequencies of the ac magnetic field to detect.
The solid lines in Fig. \ref{FigS1} show the measured amplitude $v_{1m}$ of the voltage $v_1$ as a function of the amplitude $b_{ac}$ of the ac magnetic field when the frequency of the ac magnetic field is set to $100~$Hz, $2~$kHz, $11~$kHz, $30~$kHz, and $128~$kHz, respectively.
As is shown, the voltage amplitude $v_{1m}$ increases linearly with $b_{ac}$ for small values of $b_{ac}$, which is in agreement with expectation according to Eq. \ref{P0}.
The voltage amplitude $v_{1m}$ is fit with the linear function $v_{1m}=kb_{ac}$ in the region where $b_{ac}$ is small.
The fit result is shown with the dash lines in Fig. \ref{FigS1}.
The fit value of the coefficient $k$ decreases as the frequency of the ac magnetic field increases, indicating that the approximation of quasi-static magnetic field gradually breaks down.

\begin{figure}[http]
\centering
\includegraphics[width=0.8\columnwidth]{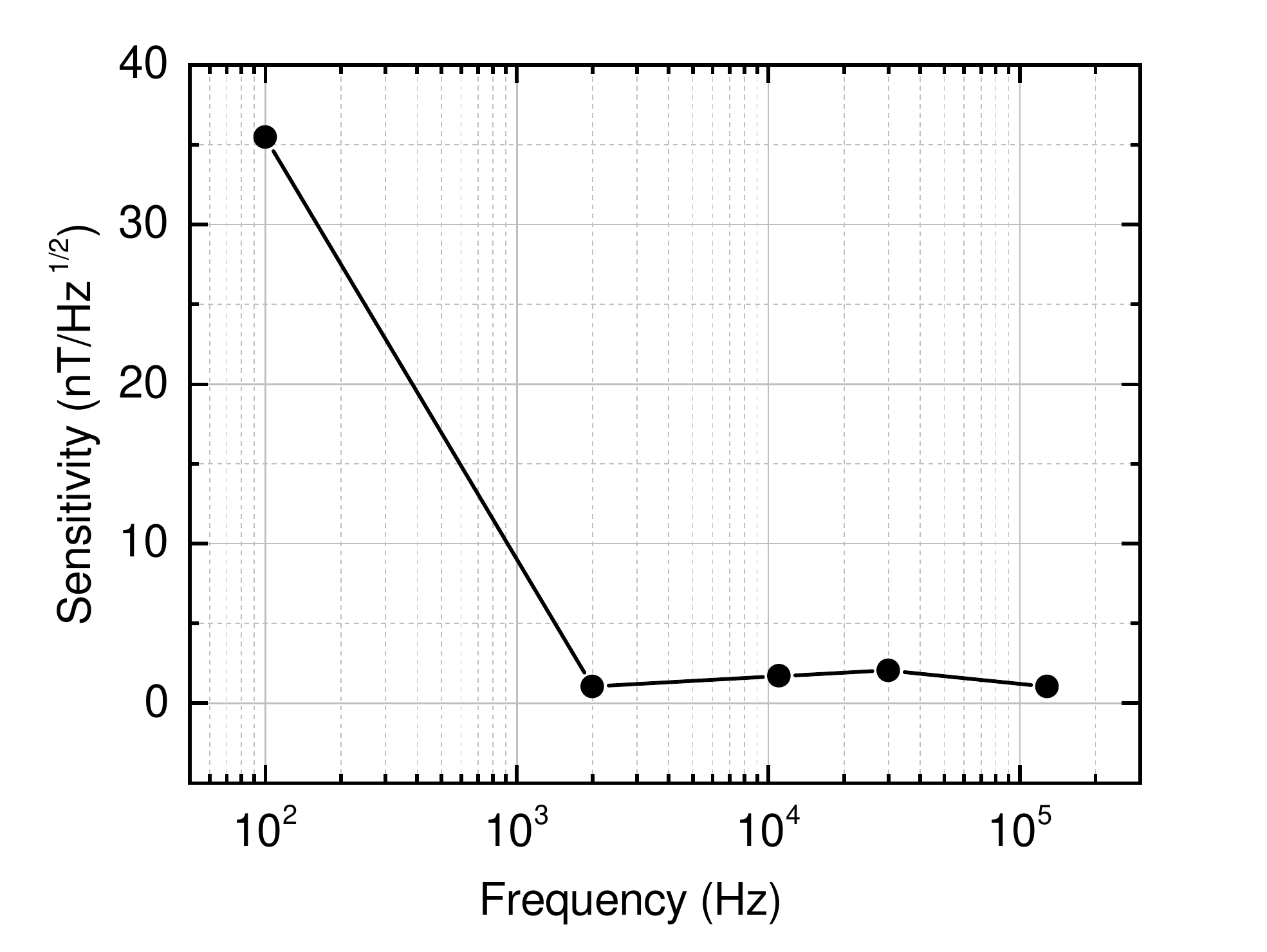}
\caption{\label{FigS2} Sensitivity of our quantum probe in the dissipative sensing protocol as a function of the frequency of the ac magnetic field to detect. The error bars in the sensitivity come from the fitting uncertainty of the coefficient $k$ when fitting the experimentally measured voltage amplitude $v_{1m}$, and are smaller than the symbol size.}
\end{figure}

The sensitivity is defined as the minimum detectable amplitude of the ac magnetic field with unit time of measurement, which depends on the noise level at the same frequency of the ac magnetic field.
The dash dot lines in Fig. \ref{FigS1} show the noise levels in a bandwidth of $1~$Hz at frequencies $100~$Hz, $2~$kHz, $11~$kHz, $30~$kHz, and $128~$kHz.
The minimum detectable amplitude $b_{ac, min}$ of the ac magnetic field, at which the signal-to-noise ratio of the measured voltage is generally considered to be 1, can be estimated as the abscissa of the intersection point of corresponding dash and dash dot lines.
The sensitivity estimated with this method is shown in Fig. \ref{FigS2}.
When the frequency of the ac magnetic field to detect is $100~$Hz, the sensitivity is about $30~$nT$/\sqrt{\textrm{Hz}}$.
As the noise level at the other measured frequencies is much smaller, the sensitivity at these frequencies improves with more than an order compared to that at $100~$Hz and approaches to $1~$nT$/\sqrt{\textrm{Hz}}$.

\begin{figure}[http]
\centering
\includegraphics[width=0.8\columnwidth]{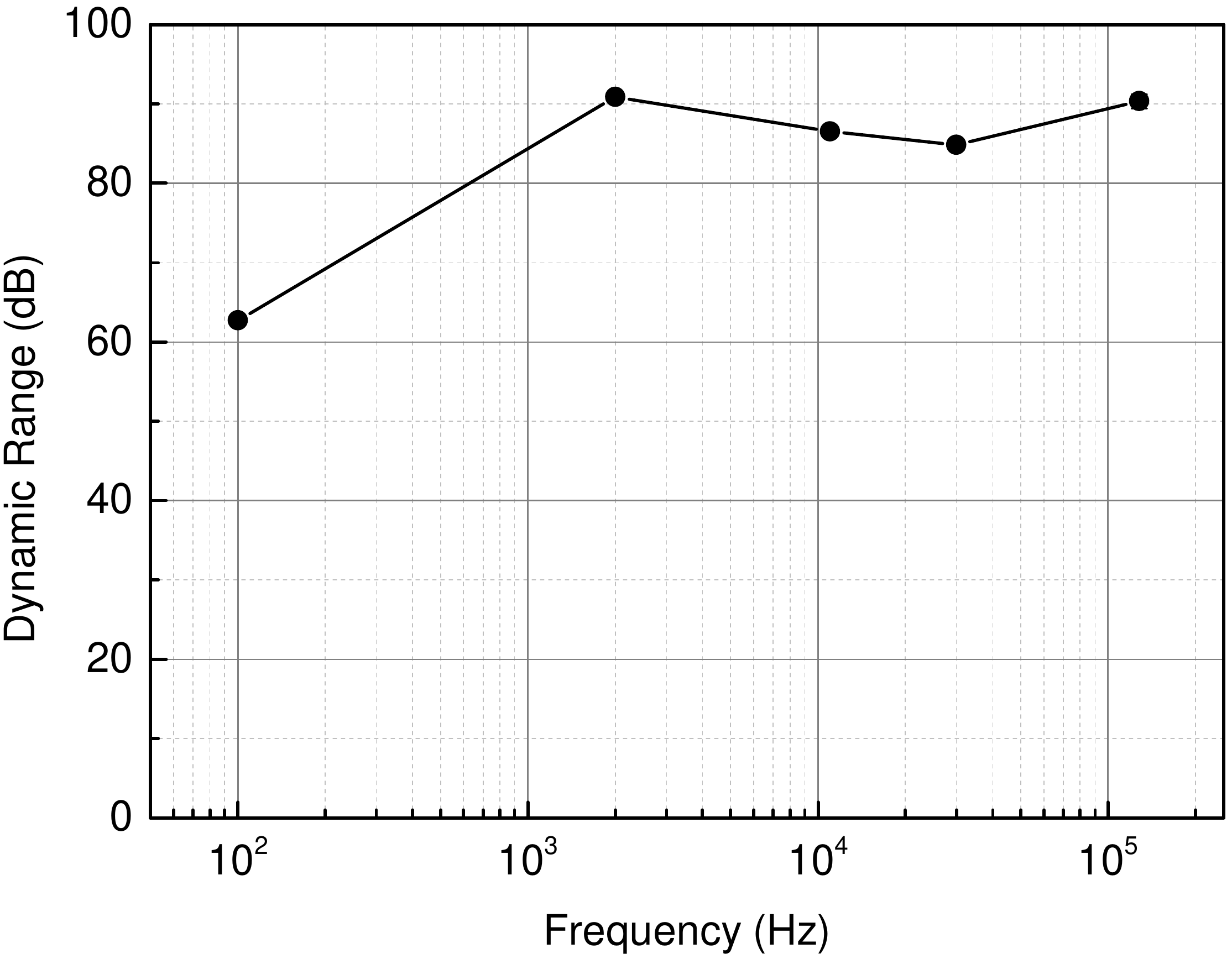}
\caption{\label{FigS3} Dynamical range of our quantum probe in the dissipative sensing protocol as a function of the frequency of the ac magnetic field to detect. The error bars in the dynamical range are mainly due to the estimation uncertainty of the maximum detectable amplitude $b_{ac}$ of the ac magnetic field, and are smaller than the symbol size.}
\end{figure}

As the amplitude $b_{ac}$ of the ac magnetic field increases to a sufficiently large value, the increase of the voltage amplitude $v_{1m}$ will be saturated.
Due to the hyperfine interaction of the quantum probe with the unpolarized $^{14}$N nuclear spin, $v_{1m}$ presents a 3-peak feature in the region of large $b_{ac}$, which is shown in Fig. \ref{FigS1}.
The maximum detectable amplitude $b_{ac, max}$ of the ac magnetic field is estimated as the abscissa of the first peak.
The short dash lines in Fig. \ref{FigS1} show the positions of $b_{ac, max}$ when the frequency of the magnetic field is $100~$Hz, $2~$kHz, $11~$kHz, $30~$kHz, and $128~$kHz.
The dynamical range is defined as $20\lg(b_{ac, max}/b_{ac, min})$.
Figure \ref{FigS3} shows the dynamical range as a function of the frequency of the ac magnetic field.
The dynamical range reaches $90~$dB when the frequency of the ac magnetic field is, e.g., $2~$kHz.

\end{document}